\documentclass[a4paper,fleqn,usenatbib]{mnras}

\usepackage{newtxtext,newtxmath}

\usepackage[T1]{fontenc}
\usepackage{ae,aecompl}

\usepackage[dvips]{graphicx}	
\usepackage{amsmath}	
\usepackage{amssymb}	
\usepackage{multicol}        
\usepackage{bm}		
\usepackage[whole]{bxcjkjatype}
\usepackage{pdflscape}	
\usepackage{color}

\title[Age Dating the Galactic Bar with the NSD]
{Age Dating the Galactic Bar with the Nuclear Stellar Disc}
\author[J. Baba \& D. Kawata]{
Junichi \textsc{Baba}$^{1}$\thanks{E-mail: jun.baba@nao.ac.jp; babajn2000@gmail.com (JB); d.kawata@ucl.ac.uk (DK)}, 
and
Daisuke \textsc{Kawata}$^{2}$
\\
$^1$ National Astronomical Observatory of Japan, Mitaka, Tokyo 181-8588, Japan\\
$^2$ Mullard Space Science Laboratory, University College London, Holmbury St. Mary, Dorking, Surrey, RH5 6NT, UK\\
}

\begin{document}

\date{Accepted 2020 January 11. Received 2019 December 26; in original form 2019 September 17}


\maketitle

\begin{abstract}
{ 
From the decades of the theoretical studies, it is well known that the formation of the bar triggers the gas funnelling into the central sub-kpc region and leads to the formation of a kinematically cold nuclear stellar disc (NSD). We demonstrate that this mechanism can be used to identify the formation epoch of the Galactic bar, using an $N$-body/hydrodynamics simulation of an isolated Milky Way-like galaxy. As shown in many previous literature, our simulation shows that the bar formation triggers an intense star formation for $\sim1$~Gyr in the central region, and forms a NSD. As a result, the oldest age limit of the NSD is relatively sharp, and the oldest population becomes similar to the age of the bar. Therefore, the age distribution of the NSD tells us the formation epoch of the bar. 
We discuss that a major challenge in measuring the age distribution of the NSD in the Milky Way is contamination from other non-negligible stellar components in the central region, such as a classical bulge component. We demonstrate that because the NSD is kinematically colder than the other stellar populations in the Galactic central region, the NSD population can be kinematically distinguished from the other stellar populations, if the 3D velocity of tracer stars are accurately measured. Hence, in addition to the line-of-sight velocities from spectroscopic surveys, the accurate measurements of the transverse velocities of stars are necessary, and hence the near-infrared space astrometry mission, {\it JASMINE}, would play a crutial role to identify the formation epoch of the Galactic bar. We also discuss that the accuracy of stellar age estimation is also crucial to measure the oldest limit of the NSD stellar population.}
\end{abstract}

\section{Introduction}

Revealing the formation history and structure of the bar in the Milky Way is a long-standing challenge in Galactic astronomy. Early infrared observations revealed that the Galactic bulge shows the boxy-shape and is believed to be a bar \citep[e.g.,][]{Nakada+1991,BlitzSpergel1991b}. This is supported by the non-circular features in the Galactic longitude and line-of-sight (LOS) velocity of the gas { in the central region \citep[e.g.,][]{Kerr1967,Peters75,MulderLiem1986,Athanassoula1988,Athanassoula1989,Binney+1991}.} Photometric and spectroscopic surveys towards the Galactic bulge, such as the Bulge Radial Velocity Assay \citep[BRAVA;][]{Kunder+2012}, the Abundances and Radial velocity Galactic Origins Survey \citep[ARGOS;][]{Freeman+2013}, VISTA Variables in the Via Lactea \citep[VVV;][]{Minniti+2010}, are revealing more detailed structure of the Galactic bar/bulge. Structure analysis using red clump stars from the VVV shows a clear inner boxy/peanut-shaped bulge connected to the long thinner Galactic bar as long as about 5~kpc \citep{WeggGerhard2013,Wegg+2015}. Furthermore, photometric data from Pan-STARRS1 \citep{Chambers+2016}, 2MASS \citep{Skrutskie+2006} and AllWISE \citep{Wright+2010}, combined with {\it Gaia} DR2 \citep{GaiaCollaboration2018} reveal the Galactic bar shape in the inner Galactic disc \citep[][]{Anders+2019}. These observations suggest that the orientation of the major axis of the Galactic bar relative to the axis along the Sun and the Galactic centre is about $25^\circ$ \citep[for reviews,][]{Bland-HawthornGerhard2016,ZoccaliValenti2016}. Using kinematic data of the bar/bulge stars, recent studies suggest that the current pattern speed of the Galactic bar is about $40~\rm km~s^{-1}~kpc^{-1}$ \citep{Portail+2017,Sanders+2019,Bovy+2019}, as opposed to a fast pattern speed inferred from the kinematics of the solar neighbourhood stars \citep{Dehnen1999b}.

Another unknown property of the Galactic bar is the formation time. The Galactic bar impacts the dynamics and star formation of the Galactic disc significantly, and identifying the formation epoch of the bar is one of the key questions to understand the formation and evolution history of the Milky Way. 
\citet{Haywood+2018} discussed that the bar formation at the early epoch quenched star formation \citep[see also][]{Khoperskov+2018}, which leads to the transition from the $\alpha$-high thick disc formation to the $\alpha$-low thin disc formation\footnote{ Note that the origin of thick and thin disc formation is still highly debated and there are various mechanisms suggested \citep[see e.g.][for a review]{KawataChiappini2016}. This is merely to provide an example related to the bar formation. In general, the varying star formation rates can explain the transition from the $\alpha$-high thick disc to the $\alpha$-low thin disc \citep[e.g.][]{Chiappini+1997,Brook+2004,Colavitti+2008}. 
}.
\citet{Friedli+1994} demonstrated that the formation of the bar induces a strong radial migration and affects the metal distribution of the disc stars \citep[see also][]{DiMatteo+2013}.
Using the distribution of infrared carbon stars, \citet{ColeWeinberg2002} suggested that the age of the Galactic bar is about 2~Gyr. In contrast, { the stellar populations of the bulge stars are found to be old \citep[e.g.][]{Ortolani+1995,Kuijken+Rich2002,Zoccali+2003} and $\alpha$-high \citep[e.g.][]{McWilliamRich1994,Zoccali+2006}.}
Recently using APOGEE \citep{Majewski+2017} and {\it Gaia}~DR2 data, \citet{Bovy+2019} analysed the age and stellar abundances of the stars within the bar region, and confirmed that the stars in the bar region dominated with older age and $\alpha$-high thick disc population. They further suspected that the Galactic bar formed when the old thick disc formed at an early epoch of the Milky Way formation, and argued that the bar age is about 10 Gyr. 

However, it should be noted that the age of stars in a bar does not equal the `dynamical' age of the bar  \citep[e.g.][]{Wozniak2007}. Since the bar is a dynamical structure formed via bar instability \citep{Hohl1971,OstrikerPeebles1973,Efstathiou+1982} or tidal interaction\footnote{Tidal interactions do not necessarily induce bar formation, but can destroy a pre-existing bar \citep[e.g.][]{Pfenniger1991,Berentzen+2003} or impede/cease bar formation \citep[e.g.][]{Moetazedian+2017}. 
} \citep{Noguchi1987,MiwaNoguchi1998} of a pre-existing disc, the stars formed prior to bar formation can be captured in the bar and the age of the stars in the bar can be older than the age of the bar itself. 
{
After a bar has fully formed, it generally experiences a buckling instability \citep[e.g.][]{Raha+1991,Combes+1990,PfennigerFriedli1991,Martinez-ValpuestaShlosman2004,Debattista+2006}. The buckling instability causes the bar to thicken out of the plane into a boxy/peanut shape \citep[e.g.][]{CombesSanders1981}, whose degree depends on kinematics of stellar populations in the pre-existing disc \citep{MerrittSellwood1994,Debattista+2017,Debattista+2020,Fragkoudi+2017}.
}

In this paper, we consider that the central sub-kpc region of the bulge is a key structure needed to infer the age of the Galactic bar. { 
Gas inflow driven by the bar formation has a long history of theoretical studies \citep[e.g.][]{SandersHuntley1976,Roberts+1979,vanAlbadaRobers1981,Shlosman+1989,Athanassoula1992b,FriedliBenz1993}. Previous hydrodynamic simulations of barred galaxies have shown that the torque of a bar induces gas inflow towards the galactic centre, resulting in the formation of a compact and kinematically cold nuclear gas disc in the central sub-kpc region of the barred galaxies \citep[e.g.][]{Athanassoula1992b,WadaHabe1992,Knapen+1995,Piner+1995,EnglmaierGerhard1997,ReganTeuben2003,ReganTeuben2004,Kim+2012a,Kim+2012b,Li+2015,SeoKim2013,Shin+2017,Sormani+2018b}.
}
It is expected that the nuclear gas disc \citep[the so-called central molecular zone, CMZ;][]{MorrisSerabyn1996} of the Milky Way galaxy developed as a direct consequence of gas inflow driven by the Galactic bar { \citep[e.g.][]{Kerr1967,Peters75,MulderLiem1986,Athanassoula1989,Ridley+2017,Sormani+2018a}}. Furthermore, $N$-body/hydrodynamic simulations of isolated barred discs showed that gas which fell into the galactic centre settled into a rotating star-forming nuclear disc { \citep[e.g.][]{FriedliBenz1995,Athanassoula2005,Wozniak2007,WozniakMichel-Dansac2009,Martel+2013,Carles+2016,Seo+2019}}. Using $N$-body/hydrodynamic simulations, \citet{Cole+2014} showed that the nuclear stellar disc (NSD) formed in a barred galaxy is thinner, younger, kinematically `cooler' and more metal rich than the surrounding stars in the bar and bulge \citep[][]{Ness+2014,Debattista+2015,Debattista+2018}. 

The NSD of the Milky Way has been indirectly inferred by modelling the infrared photometric observations with estimated density profiles \citep{Catchpole+1990,Launhardt+2002}, coinciding with the CMZ. The vertical extent of $|b| < 0.4^\circ$ (or $\sim 50$ pc) and a Galactocentric radius of $\sim 150$--$200$ pc. 
Utilising APOGEE near-infrared (NIR) spectroscopy data, \citet{Schonrich+2015} detected the rotation of the NSD. \citet{Matsunaga+2015} claimed that the line-of-sight velocities of four Cepheids are consistent with the rotation of the NSD. The NSD hosts many young massive stars \citep{Yusef-Zadeh+2009}, and there are classical Cepheids in this region \citep{Matsunaga+2011,Matsunaga+2016,Dekany+2015}, i.e. there is ongoing star formation \citep{SerabynMorris1996,vanLoon+2003,Figer+2004}. Recently using accurate photometric data, \citet{Nogueras-Lara+2019b} argued that the bulk of the NSD stars formed at least 8 Gyr ago and subsequent star formation activity was on a low level { until about 1~Gyr ago}.

{Based on the theoretical studies summarised above, the} NSD is considered to be the structure formed after the gas fell into the central disc region due to the bar formation. { Because it occupies relatively stable orbits in the bar,} 
the NSD is expected to be difficult to be disrupted unless the bar is broken somehow. {Hencce, the } age of the oldest stars in the NSD should correspond to the age of the stars which formed concurrently with the bar, from which the age of the Galactic bar can be deduced. 
{%
In fact, the VLT/MUSE TIMER Project \citep{Gadotti+2019} relies on this expectation, and observes the star formation histories in the inner disc region of external barred galaxies to identify their bar formation epoch \citep[][]{Gadotti+2015}.
}

In this paper, we perform an $N$-body/hydrodynamic simulation of an isolated Milky Way-like galaxy and test this prediction that the age of NSD tells us the formation time of the bar.
In Section \ref{sec:ModelMethod}, we describe our galaxy model and simulation method.
Section \ref{sec:NSDFormation} describes the formation and evolution of the NSD driven by the bar formation in the simulation. We analyse the age distributions of the stars in the nuclear region of the simulation in Section \ref{sec:AgeDistribution}, and discuss an observational challenge to analyse the age distribution of the NSD due to the contamination from the { underlying hot} stellar components, such as a classical bulge. Finally, we present our conclusions in Section \ref{sec:Conclusions}.

\section{Models and Methods}
\label{sec:ModelMethod}

To investigate the effect of bar formation in the Milky Way galaxy, we performed an $N$-body/hydrodynamics simulation of an isolated galactic disc. We generated the initial axisymmetric model of a Milky Way-like galaxy composed of live stellar/gaseous discs, a live classical bulge, and a fixed dark matter halo. The stellar disc follows an exponential profile with a mass of $4.3 \times 10^{10}~M_\odot$, a scale-length ($R_{\rm d}$) of $2.6$ kpc, and a scale-height of $300$ pc. Using Hernquist's method \citep[][]{Hernquist1993}, the velocity structure of the stellar disc in cylindrical coordinates was determined by a Maxwellian approximation. We set the reference radial velocity dispersion by assuming that Toomre's $Q$ at $R = 2.5 R_{\rm d}$ is equal to 1.1. The gas disc also follows an exponential profile with a total mass of $1.2 \times 10^{10}~M_\odot$, a scale-length ($R_{\rm d,g}$) of $10.4$ kpc, and a scale-height of $100$ pc. The classical bulge follows the Hernquist profile with an isotropic velocity dispersion, whose mass and scale-length are $6.7 \times 10^9~M_\odot$ and $0.79$ pc, respectively. For simplicity, we assume the dark matter halo to be a static potential, whose density profile follows the Navarro–Frenk–While (NFW) profile. We assigned the mass and concentration parameter at $1.26 \times 10^{12}~M_\odot$ and $11.2$, respectively. A more detailed model description can be found in \citet{Baba2015c}.

Note that we took into account the continuous gas accretion to the disc from the halo, by adding new gas (SPH) particle at $|z| = 5$ kpc. The gas particles were continuously added at the rate of $2~\rm M_\odot~yr^{-1}$, which is comparable to the total SFR of the simulated galaxy. The gas particles were added following the exponential radial profile with the scale-length, $R_{\rm d,g}$, within the radius of 20 kpc. 
{
The initial velocity of the added SPH particles was set to be azimuthal motion of the local circular velocity with an isotropic velocity dispersion of $10~\rm km~s^{-1}$, and the temperature to be $10^4$ K. Because the initial velocity was slightly larger than the local circular velocity because of the assumed velocity dispersion, 
the gas particles tend to fall into a larger radius than the given radius at $|z| = 5$ kpc.
}

Our simulations were performed using the $N$-body/smoothed particle hydrodynamic (SPH) simulation code {\tt ASURA-2}. To perform time integration, we used leapfrog integrator with variable and individual time steps. {\tt ASURA-2} also implements a time-step limiter \citep{SaitohMakino2009} that allows us to solve rapid expansions of the gas shell due to supernovae by imposing sufficiently small time-steps to neighbouring particles. The FAST method \citep{SaitohMakino2010}, which speeds up the time integration of a self-gravitating fluid by using different time steps for gravity and the hydrodynamic interactions of each particle, was also implemented. We computed self-gravity with the Tree/GRAPE method \citep{Makino1991} using a software emulator of GRAPE known as Phantom-GRAPE \citep{Tanikawa+2013}\footnote{https://bitbucket.org/kohji/phantom-grape.}. The simulations also take into account radiative cooling for a wide temperature range of $20~{\rm K} < T < 10^8~{\rm K}$, heating due to far-ultraviolet background radiation, probabilistic star formation from the cold dense gas, as well as thermal feedback from type II supernovae \citep{Saitoh+2008} and $\rm H_{II}$ regions \citep{Baba+2017}. This is the same model as is used in \citet{Baba+2017}, where more details of the simulations are described. Dynamics of spiral arms in the simulated barred galaxy using the same model has been presented in \citet{Baba2015c}. The dynamical evolution of the bar and bulge and link to the NSD are analysed in this paper.

\begin{figure*}
\begin{center}
\includegraphics[width=0.95\textwidth]{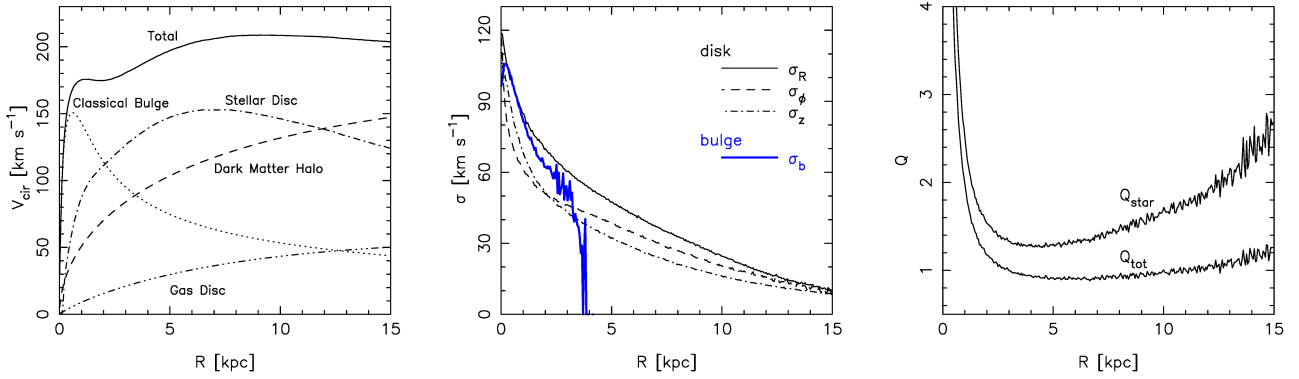}
\caption{
    Initial condition {( i.e. the system is relaxed; see Section \ref{sec:ModelMethod})} of the galaxy model. Left: circular velocity and contributions of individual components as a function of the galactocentric radius, $R$. Center: velocity dispersion of the disc stars (black curves) and radial velocity dispersion of classical bulge stars (thick solid, blue curve) as a function of $R$. Right: Toomre's $Q$ values { of the disc stars ($Q_{\rm star}$) and the disc stars+gas ($Q_{\rm tot}$)} as a function of $R$.
}	
\label{fig:init}
\end{center}
\end{figure*}

The initial numbers of stars and gas (SPH) particles are $5.7$ million and $4.5$ million respectively, 
with particle masses for star and gas particles at approximately $9.1 \times 10^3~M_{\rm \odot}$ and $3 \times 10^3 ~M_{\rm \odot}$, respectively. We used a gravitational softening length of $10$ pc, which is sufficiently small to resolve the three-dimensional structure of a disc galaxy \citep{Baba+2013}. As described above, a disc galaxy model in a near-equilibrium state was generated using Hernquist's method \citep{Hernquist1993}. Then, to allow the system to dynamically relax, it was evolved for 6 Gyr with the SPH particles fixed, and the star particles randomly displaced azimuthally at every several time steps to prevent the growth of non-axisymmetric features \citep{McMillanDehnen2007}. Next, the circular velocity of the SPH particles was reassigned based on the mass distribution after the system was relaxed. This equilibrium state was used as the `initial condition' for the numerical simulation. { 
Fig.\ref{fig:init} shows the initial circular velocity, velocity dispersion of the classical bulge and disc stars,
and initial Toomre's $Q$ value of the disc stars ($Q_{\rm star}$) as a function of the galactocentric radius, $R$. We note that gravitational interactions between the stellar and the gaseous components make a galaxy more unstable than the two components considered individually \citep{JogSolomon1984}. In the mixed component model, an effective Toomre's $Q$-value is approximately given by $Q_{\rm tot}^{-1} \approx Q_{\rm star}^{-1} + Q_{\rm gas}^{-1}$, where $Q_{\rm gas}$ is the $Q$-value of the gas \citep{WangSilk1994}. The radial profile of the initial $Q_{\rm tot}$-value is also shown in Fig.\ref{fig:init}.
 There is no vertical cut applied to select the particles. The $Q$-values are computed based on the radial velocity dispersion.}

\section{Bar-driven growth of nuclear disc}
\label{sec:NSDFormation}

\begin{figure*}
\begin{center}
\includegraphics[width=0.98\textwidth]{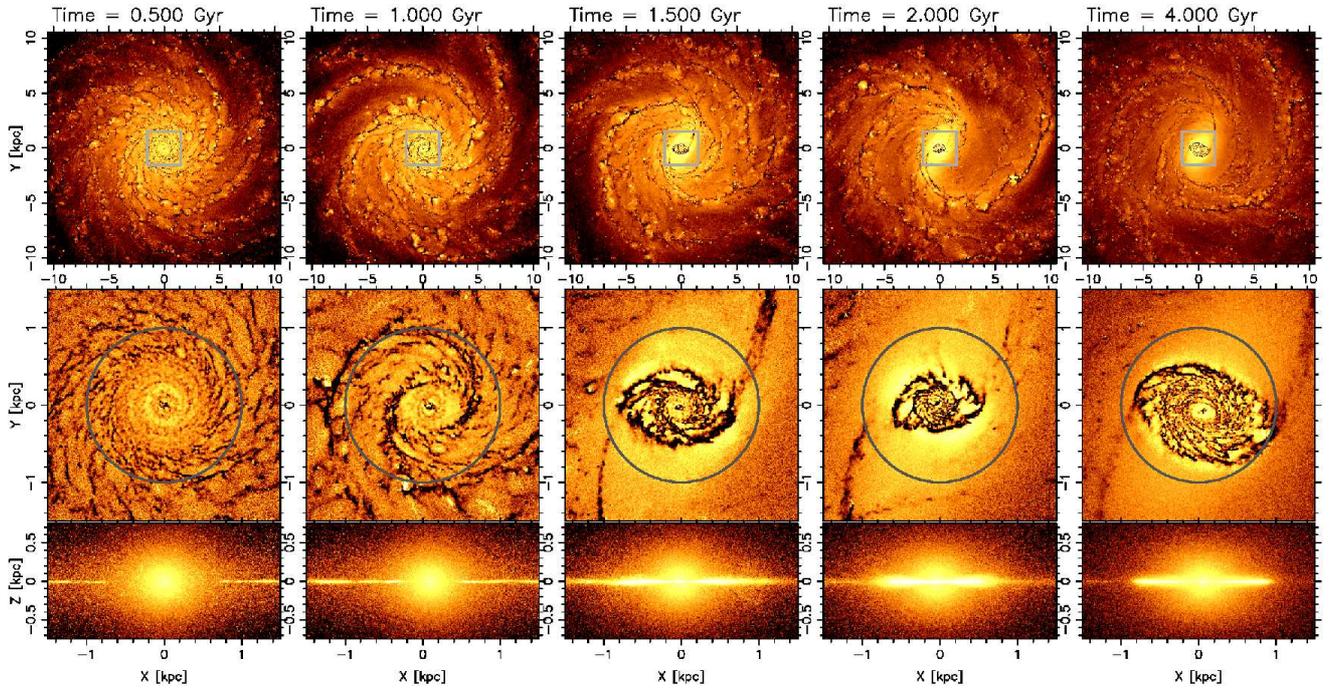}
\caption{
	Morphological evolution of the simulated barred spiral galaxy. Colors indicate surface density of stars in logarithmic scale. Top: Evolution of face-on view ($x$-$y$) of the whole galaxy scale. The gas (mainly in molecular gas) are shown in the dark filamentary structure. Middle: Evolution of face-on views of the central region (corresponding to the squared regions in the top panels). After the bar formed ($t \gtrsim 1.5$ Gyr), the major-axis of the bar is set to be the direction to $25^\circ$ from the $y$-axis. The regions analysed in Fig.~\ref{fig:BarEvolution} are enclosed by circles with a radius of 1 kpc. { See also Appendix \ref{sec:gasring} for the evolution of the nuclear gas disc.} Bottom: Evolution of edge-on views of the central region.
}	
\label{fig:SnapshotEvolution}
\end{center}
\end{figure*}

\begin{figure}
\begin{center}
\includegraphics[width=0.45\textwidth]{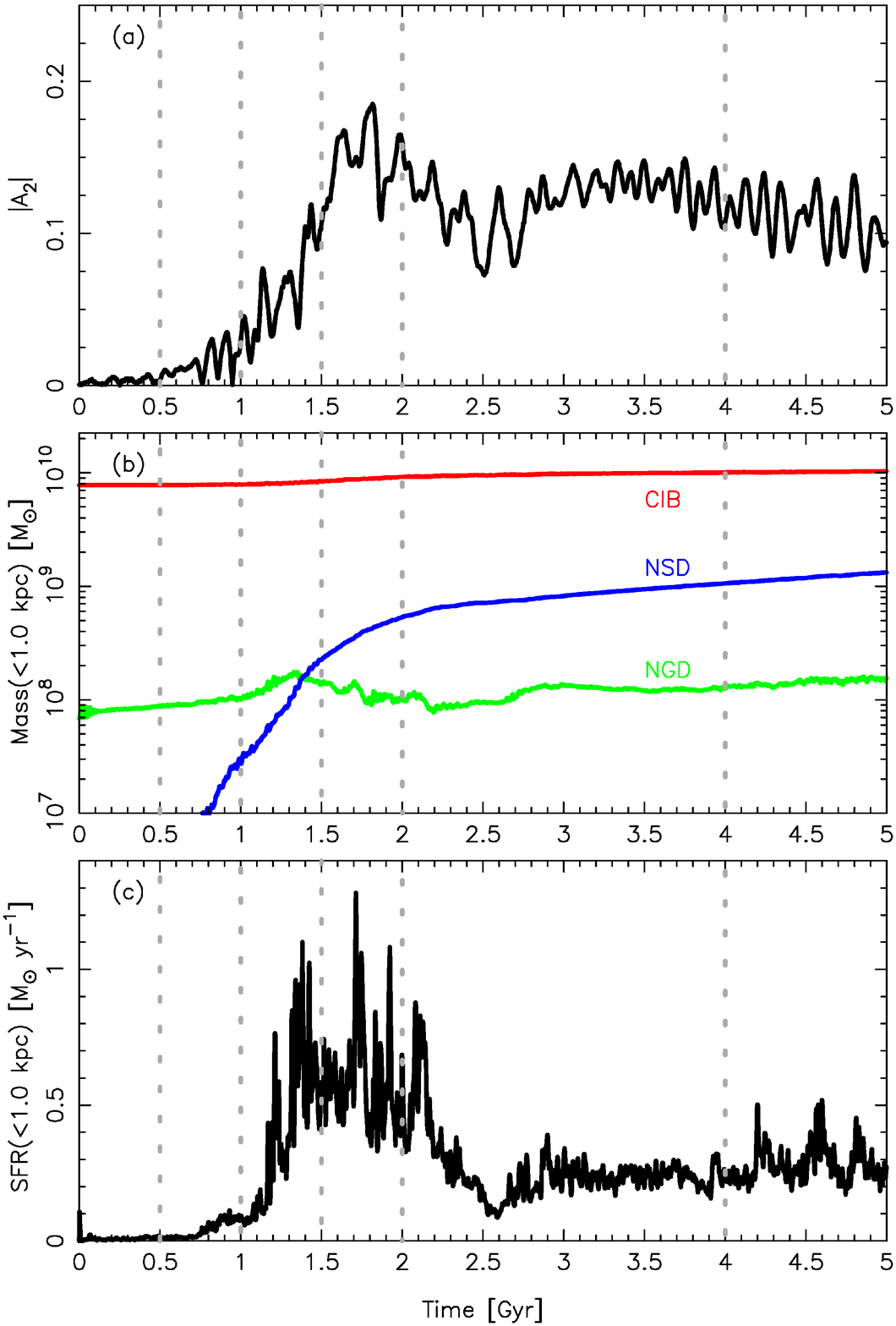}
\caption{
	Time evolution of (a) the bar amplitude within $R = 3.5$ kpc, (b) the masses and (c) in-situ SFR in the central 1 kpc region as highlighted in the middle rows of Fig.\ref{fig:SnapshotEvolution}. In panel (b), components labelled by `ClB`, `NSD' and `NGD' are defined as the old stars (i.e. classical bulge and old disc stars), newly born stars, and gas in this region, respectively. Vertical dashed lines indicate the times corresponding to those of the snapshots shown in Fig.\ref{fig:SnapshotEvolution}.
}	
\label{fig:BarEvolution}
\end{center}
\end{figure}

We run the simulations for about 5 Gyr and display time evolution of face-on stellar distributions in the top row of Fig. \ref{fig:SnapshotEvolution}. In { this particular} model, grand-design spiral arms form in the discs at $t \lesssim 1$ Gyr. A stellar bar starts developing around $t = 1.0$ Gyr, and a bar with a semi-major axis of $\sim 3$ kpc is fully developed at $t=1.5$ Gyr. { These time-scales of the spiral and bar formation are due to the initially unstable initial condition.}
To investigate time evolution of the stellar bar, we measure the bar strength and pattern speed with the $m=2$ Fourier amplitude of the stellar surface mass density, 
$A_2$, given by 
\begin{equation}
 A_2 = \frac{\sum_{j=1}^{N} m_j e^{2i\phi_j}}{\sum_{j=1}^{N} m_j}.
\end{equation}
Here $m_j$, $\phi_j$, and $N$ are the mass and azimuth angle of a $j$-th stellar particle, and the number of stellar particles within a cut-off radius of $R_c = 3.5$ kpc, respectively \citep[e.g.][]{Dubinski+2009}.
Fig. \ref{fig:BarEvolution}(a) shows the evolution of the bar amplitude ($|A_2|$). The bar starts to grow in a time of $t \simeq 1$ Gyr, and then reaches the maximum amplitude around $t \simeq 1.8$ Gyr. After that, the bar slowly decreases their amplitude.

The middle and bottom panels of Fig. \ref{fig:SnapshotEvolution} respectively show the $x$-$y$ and $x$-$z$ maps of the stars ({ coloured} by orange) and gas (dark) distribution in the central region (enclosed by squares in the top panels). We can see the stellar and gas discs in the central sub-kpc region in the snapshots after the bar formation starting at $t=1$~Gyr. Hereafter, we refer these stellar and gas disc structures as NSD and nuclear gas disc (NGD), respectively. The radius and thickness of the NSD are $\sim 800$ pc and $< 100$ pc, respectively. The radius and thickness of the NGD are $\sim 800$ pc and $\lesssim 10$ pc, respectively { (See also Appendix \ref{sec:gasring}).} These sizes are much larger than the observed NSD in the Milky Way. For example, the radial extent of the NSD in the Milky Way is around 230~pc \citep[][]{Launhardt+2002}, and the vertical scale-height is measured to be around 45~pc \citep[][]{Nishiyama+2013}. Note that the aim of this study is to explore the phenomenological link between the Galactic bar structure and the NSD, not to reproduce the NSD and NGD structures in the Milky Way with the numerical simulation. Therefore, the discrepancy in the size of the nuclear discs between our simulation and the Milky Way is not an issue in this study. We consider that the discussion in this paper does not depend on the size of the nuclear disc.

Fig. \ref{fig:BarEvolution}(b) shows the time evolution of masses of the gas and stars in the central 1 kpc region. We consider that the gas and the new born stars after $t=0$~Gyr reside mainly in the nuclear disc in our simulation. Hence, we refer to the gas and the new born stellar mass within 1 kpc as NGD and NSD in this panel. 
After bar formation starts at $t \gtrsim 1$ Gyr, the NSD mass ($M_{\rm NSD}$) rapidly increases until the bar is fully formed at $t\simeq1.5$~Gyr. At this point, the increase of the NSD mass slow down, reaching $10^9~\rm M_\odot$ at $t = 3$ Gyr. This mass is similar to the NSD mass of the Milky Way, around $1.4\times10^9$~M$_{\odot}$ \citep{Launhardt+2002}. In contrast, the NGD mass ($M_{\rm NGD}$) increased as the bar grew at $t=1-1.5$~Gyr due to the inflow of gas triggered by bar formation (see also Appendix \ref{sec:gasdepletion}). However, the NGD mass reached a quasi-steady value of about $10^8~\rm M_\odot$ after the bar fully formed. This mass is also similar to the observed value for the NGD of the Milky Way, around $5 \times 10^7$~M$_{\odot}$ \citep{Launhardt+2002}, although the size of the disc is much larger. Note that the bar formation triggers inflow of not only gas but also older stars into the central region. In fact, $M_{\rm ClB}$ in Fig. \ref{fig:BarEvolution}(b) represents the evolution of the mass of the old stars, i.e. classical bulge and old disc stars, already in placed at $t=0$~Gyr in our simulation; it increases by a factor of about 1.2 after the bar formation.

The decrease of $M_{\rm NGD}$ after $t \simeq 1.3$ Gyr is mainly due to the consumption of gas during star formation. As shown in Fig. \ref{fig:BarEvolution}(c), in-situ star formation rates (SFR) in the central 1 kpc region rapidly increased at $t = 1.2$ Gyr, reaching a maximum value in the growth phase of the bar ($t \sim 1.0-1.5$ Gyr). Then, the SFR decreases to a value of about $0.25~\rm M_\odot~yr^{-1}$ with some intermittent spikes. This quasi-steady value is roughly consistent with the values obtained by the previous simulations \citep[e.g.][]{KimSaitoh+2011,Shin+2017} and the observations of the Milky Way of $\simeq0.01-0.1\rm M_\odot~yr^{-1}$ \citep[e.g.][]{Yusef-Zadeh+2009}.

Interestingly, starburst occurs during the bar growing phase ($1~{\rm Gyr}< t < 2~{\rm Gyr}$). Previous hydrodynamics simulations of barred galaxies have shown that gas inflow along the bar usually induces the increased star formation in the central { sub-kpc} regions { \citep[e.g.][]{HellerShlosman1994,FriedliBenz1995,Seo+2019}} with an associated reduction of star formation in the bar \citep{Fanali+2015,Donohoe-Keyes+2019} (See Appendix~\ref{sec:gasdepletion} for the evolution of the gas density profile due to the bar formation).
This result suggests that revealing the star formation history of the NSD in the Milky Way galaxy can be used to help identify the formation epoch of the Galactic bar, because the NSD consists of the stars formed after the bar formation and the age distribution of the NSD should be peaked at the age of the bar formation. The definition of the bar formation epoch is not clear, as shown in these results it spreads over from $t=1.0$ to $t=1.8$~Gyr in this simulation. Just for a convenience of the discussion below, in this paper we consider $t=1.5$~Gyr as the formation time of the bar in this simulation, because morphologically the bar is fully developed (Fig.~\ref{fig:SnapshotEvolution}), and it is about a middle of the bar formation period. However, this is merely a rough time of reference for the formation time of the bar. It rather means that the bar formation period is around this time and spreads over about 1~Gyr.

\section{Age distribution of Nuclear Disc Stars}
\label{sec:AgeDistribution}

As shown in the previous section, gas inflow due to bar formation causes the subsequent intense star formation.
Hence, it can be estimated when the bar is formed from the age distribution of the NSD. 
In practice, it is challenging to estimate the age of a star precisely, however, there are various methods to infer the stellar age for different stellar populations \citep[see][for a review]{Soderblom2010}. In this theoretical work, first we consider an ideal case that the age of the tracer stellar population is accurately measured in some way, and the effects of the observational uncertainties are briefly discussed later in Section \ref{sec:Conclusions}. A remaining challenge is to measure the age distribution of the NSD accurately, when there are contamination from the other stellar components in the Galactic centre region. 

One of still unknown stellar component, which could be a { non-negligible `hot'} stellar component in the central region of the Milky Way galaxy, is a `classical bulge', which may be formed in the early stages of galaxy formation \citep[][]{KormendyKennicutt2004}, and hidden in the current observational constraints \citep[][]{Shen+2010,Kunder+2016}. In fact, as shown in Section~\ref{sec:ModelMethod}, our simulation includes a classical bulge component.
In this section, we discuss how the age distribution of the NSD can be recovered if there is a non-negligible classical bulge component in the central region of the Milky Way. The exercise below does not intend to evaluate feasibility of identifying the age distribution of the NSD in any particular observational data, but to demonstrate what kind of the observational information would be required to minimise the contamination from the other { hot} stellar component, and extract the age distribution of the NSD only.

To this end, we placed an observer at a distance of $8$ kpc from the galactic centre in the disc mid-plane of the simulated disc galaxy with an angle of $25^\circ$ from the major axis of the bar \citep{Bland-HawthornGerhard2016}.
Because we are interested in the relative velocity within the central region, we do not consider the uncertainty of the solar peculiar motion either, that is, the observer is assumed to have a rotational speed of~$200~\rm km~s^{-1}$, and no vertical or radial velocity for simplicity.
Again, we consider an ideal case that one star particle is one tracer star in the central region, and ignore any observational error. 

\begin{figure*}
\begin{center}
\includegraphics[width=0.98\textwidth]{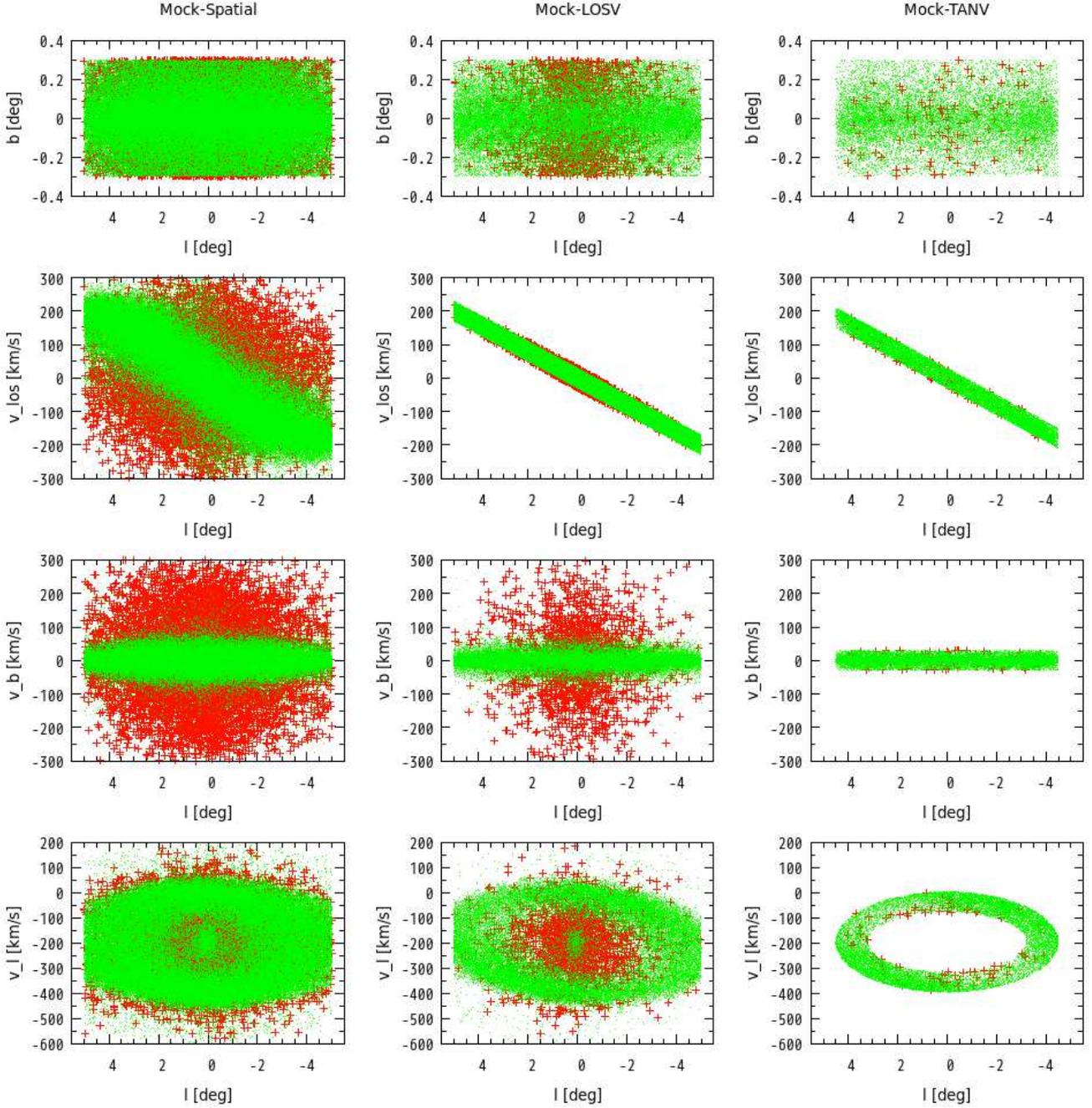}
\caption{
	Galactic longitude $l$--latitude $b$ (1st row), $l$--LOS velocity $v_{\rm los}$ (2nd row), $l$--latitudinal transverse velocity $v_b$ (3rd row), and $l$--longitudinal transverse velocity $v_l$ distributions (4th row) of star particles selected from the central region in the simulation. Crosses (red) and dots (green) indicate 
	the classical bulge stars and NSD stars, respectively. 
	From left to right columns, the samples labelled by Mock-Spatial, Mock-LOSV, and Mock-TANV are presented, respectively (see Section \ref{sec:AgeDistribution} for details). 
}	
\label{fig:NBStarSelection}
\end{center}
\end{figure*}

We first select the star particles within the volume of Galactic longitude $|l|< 5^\circ$, Galactic latitude $|b|<0.3^\circ$, and distance from the observer of $7 < d <9$ kpc\footnote{ 
This distance cut has been made for simplicity. This can be achieved if the uncertainty in the distance measurement is less than 10~\%, which could become feasible with the future NIR astrometry missions, such as {\it JASMINE} and {\it GaiaNIR} (see Section~\ref{sec:Conclusions}).
}, to spatially extract the stars in the NSD of the simulated galaxy. We name this sample of stars the `Mock-Spatial' sample. Left panels of Fig.~\ref{fig:NBStarSelection} show the spatial distribution in Galactic coordinate, $l$--$v_{\rm los}$ distribution, $l$--$v_b$ distribution, and $l$--$v_l$ distribution from top to bottom for the Mock-Spatial sample, where $v_{\rm los}$, $v_b$ and $v_l$ are line-of-sight (LOS) velocity, and transverse velocities in the directions of Galactic latitude and longitude, respectively. In Fig.~\ref{fig:NBStarSelection}, green dots represent NSD stars, which are defined as new born stars formed after the simulation started, because the majority of the new born stars formed in the NSD in the central region (Fig.~\ref{fig:BarEvolution}). In contrast, red crosses represent the classical bulge component, which was initially placed in the simulation. The Galactic longitude and latitude selections for the Mock-Spatial sample were made to focus on the region where the NSD is prominent, however, significant contamination from the classical bulge component is evident. 
In this sample, we found the classical bulge particles with $8.37 \times 10^8~ \rm M_\odot$ and the NSD particles with $8.31 \times 10^8~ \rm M_\odot$ (Table \ref{tab:Statictics}).
Note that in the central region, there is also a stellar disc component, which represents both thick and thin discs. However, the age distribution of the thin disc is spread over a wide range, and it is easier to be distinguished from the NSD age distribution, which should suddenly increase when the bar formed. The thick disc could have a peaked age distribution, which could blur the sudden increase of the NSD stars at the formation epoch of the bar, if their formation epoch is closer. However, in our simulation, the age distribution for an old thick disc is similar to what we consider for the classical bulge below. Also, in the central region, velocity dispersion of the thick disc should be quite high, and kinematically similar to the classical bulge component. Therefore, in this paper we only consider the contamination from the classical bulge component, which we believe would be most serious in the central region.

\begin{figure*}
\begin{center}
\includegraphics[width=0.98\textwidth]{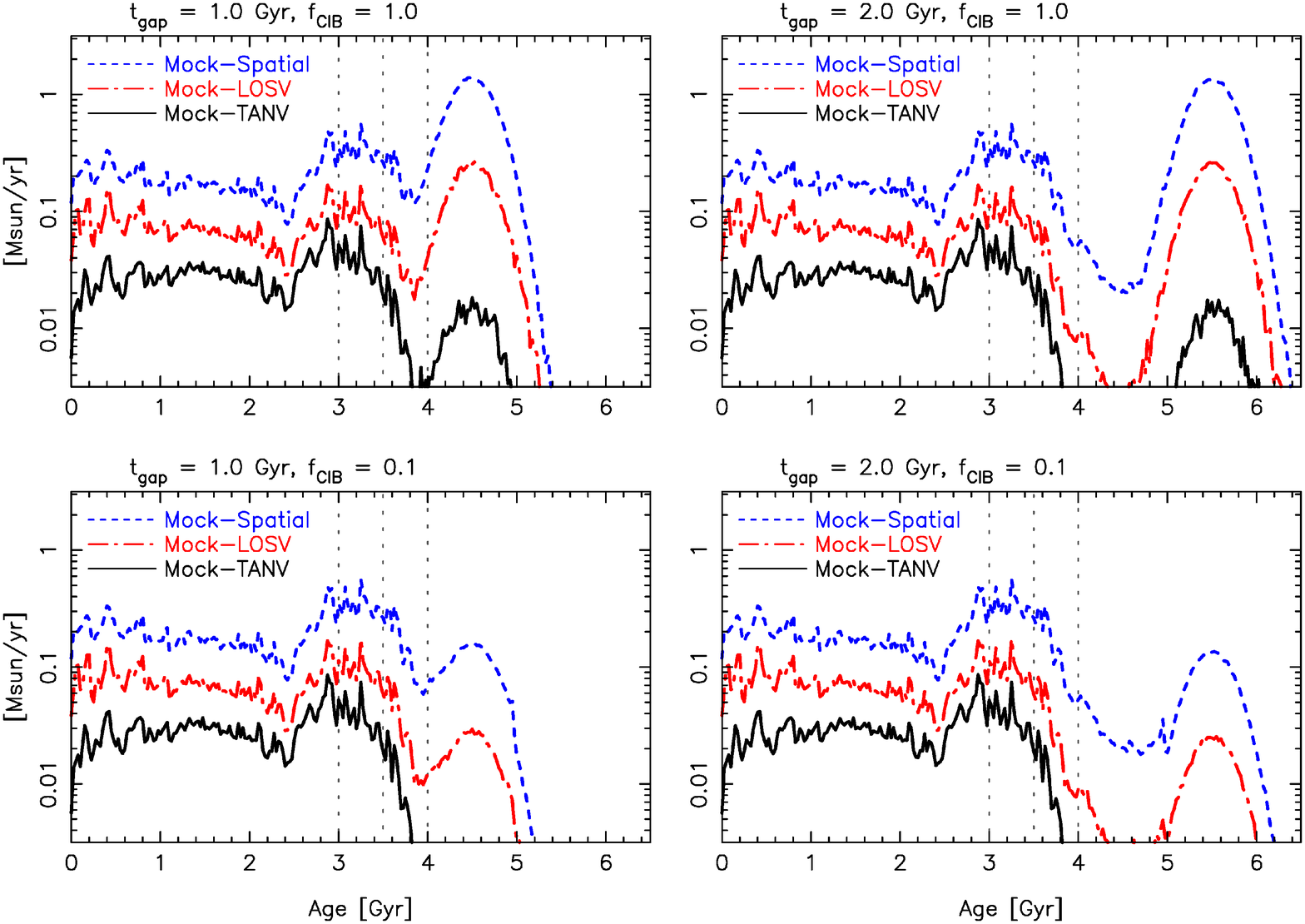}
\caption{
	Upper: Age (i.e. look-back time from $t = 5$ Gyr) distributions of the stars in the Mock-Spatial (blue dashed lines), Mock-Spatial (red dot-dashed lines) and Mock-TANV (black solid lines) sample. The age of the classical bulge is assumed to be $1$ and $2$ Gyr older than the age of the bar, Age$_{\rm bar}=3.5$~Gyr, from left to right, respectively. Vertical dashed lines correspond to $t = 1$, $1.5$, and $2.0$ Gyr from Fig.\ref{fig:SnapshotEvolution} from the right to left, respectively.
	Lower: Same as the upper panels, but for the case where one tenth ($f_{\rm ClB} = 0.1$) of the classical bulge mass is used in the simulation.
}	
\label{fig:AgeDistribution}
\end{center}
\end{figure*}

The age distribution of the Mock-Spatial sample is shown in top panels of Fig.\ref{fig:AgeDistribution} with a blue dashed line. Here, we assumed that the age distribution of the classical bulge follows a Gaussian distribution with a mean age, which is referred to as the formation time, and a dispersion of 0.25 Gyr, to mimic a starburst of classical bulge. 
We define $t_{\rm gap}$ as the time difference between the classical bulge formation time and the bar formation time, which corresponds to Age$_{\rm bar}=3.5$~Gyr (or $t=1.5$~Gyr in Fig.~\ref{fig:BarEvolution}). Top panels of Fig.~\ref{fig:AgeDistribution} show the results with different $t_{\rm gap}$ obtained by adjusting the formation time of the classical bulge, but with a fixed age of bar formation. When $t_{\rm gap}$ is greater than 2 Gyr, the formation of the NSD can be clearly distinguished from the starburst of classical bulge formation. As a result, a clear drop of stars older than Age$_{\rm bar}=3.5$~Gyr can be easily identified. However, when $t_{\rm gap}=1$~Gyr (top-left panel in Fig.~\ref{fig:AgeDistribution}), it is difficult to identify the oldest age of the NSD, as it overlaps with the age distribution of the classical bulge, and the Mock-Spatial sample contains significant classical bulge components. Note that the quantitative discussions provided here do not precisely apply to the Milky Way, because we do not know the mass or age distribution of its classical bulge or NSD. For example, if the distribution of the age of the classical bulge is much broader, a larger $t_{\rm gap}$ would be required to distinguish between the NSD from the classical bulge. However, if the mass or number of the tracer stars of the classical bulge is smaller, it would be easier to identify the NSD formation time. To demonstrate this, the lower panels of Fig.~\ref{fig:AgeDistribution} show the results if the contribution from the classical bulge is one tenth of the assumed value in the upper panel. The contribution of the classical bulge is significantly reduced; however, it is still challenging to distinguish the gap of the age between the classical bulge and the NSD, if the age gap is too small. 

We can use velocity information to further constrain the sample selection of the NSD to reduce contamination from the classical bulge { (i.e. hot component)}. LOS velocity in the Galactic centre is already obtained with the current facility \citep[e.g.][]{Matsunaga+2015,Schonrich+2015}. We therefore add the selection criterion using the LOS velocity information ($l-v_{\rm los}$) to Mock-Spatial. The middle panels of Fig.\ref{fig:NBStarSelection} show this sample of `Mock-LOSV', and the second top panel shows our selection using the LOS velocity. In general, the classical bulge is not rotating, having more isotropic velocities, whereas the NSD is rotation dominant. Therefore, the classical bulge component contamination is reduced by the LOS velocity selection, and the NSD component is relatively increased (see Table \ref{tab:Statictics}).
The age distribution of the Mock-LOSV sample is shown by the red dot-dashed line in Fig.\ref{fig:AgeDistribution}. Although the contamination of the classical bulge component is less than that of the Mock-Spatial sample, it can be seen that it is still not easy to distinguish the NSD age distribution from the age distribution of the classical bulge component for the case of $t_{\rm gap} = 1$ Gyr. Even if the classical bulge mass is reduced to one tenth of what used in the simulation (lower panels), it is difficult to distinguish the NSD from the classical bulge component with $t_{\rm gap} = 1$ Gyr.

Finally, we consider the case where transverse velocity information is available, and select the sample using the full 3D velocity information. Since the NSD stars are kinematically colder, $|v_b|$ should be small, which is seen in the 3rd row panels in Fig.~\ref{fig:NBStarSelection}. The NSD stars also are distributed in a ring-like structure in the $l$--$v_l$ plane, as seen in the bottom panels in Fig.~\ref{fig:NBStarSelection}, because of the significant rotation of the NSD. Based on these, we further reduce the sample from the Mock-LOSV sample (see Table \ref{tab:Statictics}), using $|v_b|$ and $v_l$ as shown in the 3rd and 4th row panels in the right column of Fig.~\ref{fig:NBStarSelection}, which we refer to as the `Mock-TANV' sample. The age distribution of the Mock-TANV sample is shown by the black solid line in Fig.~\ref{fig:AgeDistribution}, where it can be seen that contamination by the classical bulge component is significantly reduced. Consequently, the NSD stands out in this sample, and it is possible to identify the formation time of the NSD, i.e. the bar formation time.

These results demonstrate that in order to identify the formation time of the NSD, the challenge to reduce the contamination from the classical bulge must be addressed due to their spatial overlap. Our results highlight that obtaining 3D velocity information of the tracer sample is an effective way to reduce the contamination and extract the NSD component more clearly.

\begin{table}
\begin{center}
\begin{tabular}{ccc} \hline
 Mock           & ClB   & NSD \\
  \hline
 Mock-Spatial   & 8.4   & 8.3\\
 Mock-LOSV      & 1.6   & 2.8\\
 Mock-TANV      & 0.1   & 1.1\\
  \hline
\end{tabular}
\caption{
    The stellar mass in the nuclear stellar disc (NSD) and the classical bulge (ClB) components, when the different selections are applied, as shown in Fig.\ref{fig:NBStarSelection}.
    The unit is $10^8~M_\odot$.     
}
\label{tab:Statictics}
\end{center}
\end{table}

\section{Discussion}
\label{sec:Conclusions}

Using an $N$-body/SPH simulation of a Milky Way-like galactic disc, we demonstrate that the NSD forms from the excessive gas falling into the central region when the bar forms, and hence, a sudden drop in the number of the old stars in the age distribution of the NSD reveals the formation time of the bar. Bar formation triggers the inflow of the gas into the central region, which causes a rapid increase of star formation in the nuclear disc region. Once the NSD forms, stars are stable in the bar; hence, the entire stellar population should remain in the NSD as long as the bar exists. As a result, the rapid increase of the star formation at the formation epoch of the NSD is readily identified as the oldest edge of the age distribution within the current NSD, which tells us the formation epoch of the bar \citep[see also][]{Gadotti+2015,Gadotti+2019}. { The mechanism of the formation of the NSD has been well studied over several decades. Our study sheds a new light on the well known mechanism, and we demonstrate that it can be used to infer the formation epoch of the Galactic bar.}

We note that the model galaxy in this study is an { idealized} isolated disc model. Thus, the resulting NSD is purely of the internal origin. If the galactic disc experienced a merger, the bar and NSD could be destroyed \citep{Sarzi+2015}. Such an external effect on the NSD is beyond the scope of this paper, but would be interesting to explore with cosmological simulations \citep[e.g.][]{Buck+2018,Debattista+2019,Fragkoudi+2020}. For the Milky Way, it is considered that the last major merger occurred about 10 Gyr ago \citep{Helmi+2018,Belokurov+2018}, following which the Galactic disc experienced rather quiet evolution \citep{Brook+2004}. As such, it is likely that the bar in the Milky Way has not been disrupted since formation. Thus, the NSD is likely to have survived since formation of the bar, which allows us to use the stellar age distribution of the NSD to provide a robust way of measuring the age of the Galactic bar. { If strong feedback from the central black hole \citep[e.g.][]{Shlosman+1989} destroyed the NGD and suppressed the star formation of the NSD for some period, the age distribution of the NSD would not be smooth. However, as long as the compact and kinematically cold NSD is not destroyed, the oldest population should stay in the NSD, which still helps us to infer the epoch of the Galactic bar formation. 

We do not mean to claim that the age of the NSD is the only one or best way to identify the age of the bar. Ideally, we should have several independent ways to infer the epoch of the bar formation, so that the bar formation epoch would be more confidently measured. More studies of the bar formation and their impact to the stellar population distribution in the Milky Way are encouraged.}

A challenge in identifying the stellar population in the NSD is to distinguish it from the { underlying `hot'} stellar component, such as a classical bulge, whose mass in the central region is still unknown but could be non-negligible. We demonstrated that 3D velocity information is crucial to minimise the contamination of a classical bulge component in order to clearly identify the formation time of the NSD. LOS velocities can be obtained from the NIR multi-object spectrographs, such as APOGEE-2 \citep{Blanton+2017} and MOONS at the ESO/VLT \citep{Cirasuolo+2016}. Transverse velocities are required to be measured with astrometry. Unfortunately, the optical astrometry mission, {\it Gaia} \citep{GaiaCollaboration2016}, cannot see the NSD because of heavy dust extinction in the optical band. Recently, the proper motion was measured from the VVV survey data and the VIRAC catalogue \citep{Smith+2018}, with a median uncertainty of 0.67~mas~yr$^{-1}$ for stars with $11<Ks<14$~mag. The absolute proper motion is also measured using the {\it Gaia} reference frame \citep{Clarke+2019,Sanders+2019a}.

\begin{figure*}
\begin{center}
\includegraphics[width=0.98\textwidth]{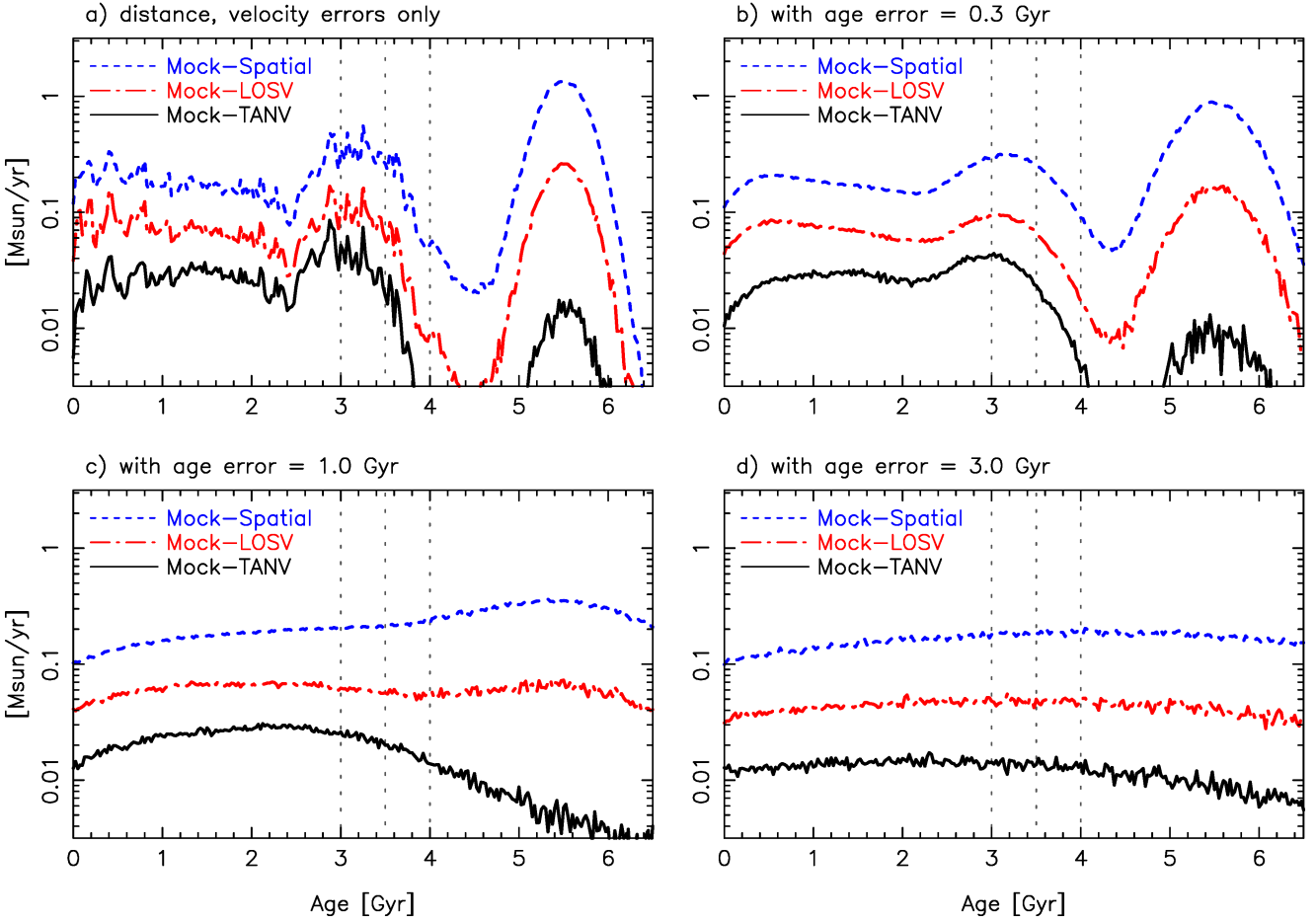}
\caption{
	Same as the upper right panel ($t_{\rm gap} = 2$ Gyr, $f_{\rm ClB} =1$) of Fig.\ref{fig:AgeDistribution}, but including the Gaussian observational uncertainties for distance, velocity, and stellar age. a) The distance and velocity uncertainties of $\sigma_{\rm d} = 1.6~\rm kpc$, $\sigma_{v_l,v_b} = 1~\rm  km~s^{-1}$ and $\sigma_{v_{\rm los}} = 0.5~\rm km~s^{-1}$ are added. b) An additional age uncertainties of 300 Myr (c), 1~Gyr (b) and 3~Gyr (c) are assumed. 
}	
\label{fig:AgeDistributionError}
\end{center}
\end{figure*}

Ultimately, NIR astrometric space missions, such as the {\it Japan Astrometry Satellite Mission for INfrared Exploration} \citep[{{\it JASMINE};}][]{Gouda2012}\footnote{
http://jasmine.nao.ac.jp/index-en.html
} and {\it GaiaNIR} \citep{Hobbs+2016,Hobbs+2019}, will provide the accurate measurement of the transverse velocity of the NSD stars, which enables to identify the age of NSD clearly and hence a definitive formation time of the Galactic bar. {\it JASMINE} is planned to be launched in mid-2020s, and is designed to achieve {\it Gaia}-level astrometric accuracy at the Galactic centre in the NIR band ($H_w$-band, 1.1$\sim$1.7~$\mu$m). {\it JASMINE} will observe the Galactic centre region within about 200~pc from the Galactic centre, and will achieve the parallax accuracy of $\sigma_{\pi}\approx 25$~$\mu$as and the proper motion accuracy of $\sigma_{\mu}\approx 25$~$\mu$as~yr$^{-1}$ for the objects brighter than $H_w=12.5$~mag and the proper motion accuracy of $\sigma_{\mu}\approx 125$~$\mu$as~yr$^{-1}$ for the objects brighter than $H_w\approx 15.0$~mag.
Hence, {\it JASMINE} will achieve about $1.6~\rm kpc$ and $1~\rm km~s^{-1}$ accuracy in distance and transverse velocities ($v_l$ and $v_b$), respectively, for bright tracer stars in the Galactic centre region, respectively. Fig.\ref{fig:AgeDistributionError}(a) shows the age distributions of the stars in the central region (as shown in the top-right panel of Fig.~\ref{fig:AgeDistribution} with $t_{\rm gap}=2$~Gyr) with these uncertainties included. We can see that the distance and velocity uncertainties minimally affect the selection of the stars and therefore minimally affect the age distribution\footnote{
Additional complexity in real observational data is crowding. However, because {\it JASMINE} focuses on only brighter stars, crowding is less serious.
}.

However, the accuracy of age estimation is crucial. To assess the effect of the age uncertainty, we add the age uncertainties of $\sigma_{\rm age} = 0.3$, 1, and 3 Gyr in Figs.~\ref{fig:AgeDistributionError}(b), (c) and (d), respectively. If the age uncertainty is as small as 0.3 Gyr, it does not affect the age distribution analysis significantly (Fig.\ref{fig:AgeDistributionError}b). Age uncertainty greater than 1 Gyr (Figs.~\ref{fig:AgeDistributionError}c and d) makes it difficult to identify the sharp decrease of the NSD stars, because the time scale of the bar formation is about 1~Gyr. Then, age uncertainty greater than 1 Gyr also makes it difficult to identify the gap between the NSD and the classical bulge stellar population, depending on the uncertainty and the difference in their ages. This highlights that measuring the age of stars in the Galactic centre region is crucial to age dating the Galactic bar through the age of the NSD stars. For example, Mira variables are bright enough to be observed with {\it JASMINE} \citep{Matsunaga+2009}, and are also known to follow the age-period relation \citep{Feast+2006,Grady+2019}. Hence, the accurate calibration for the age measurement of Mira variables is crucial for {\it JASMINE} to identify the formation time of the NSD; the formation time of the Galactic bar can then be deduced from their age distribution.

\section*{Acknowledgements}
We thank Jason Sanders and Noriyuki Matsunaga for useful discussions on the observed properties of the Galactic nuclear bulge and variable stars. We also thank Takayuki R. Saitoh for technical supports on performing numerical simulations with {\tt ASURA-2}. We also thank Dimitri Gaddotti for making us aware of the TIMER survey. We are also grateful to the referee for valuable comments. Calculations, numerical analyses and visualization were carried out on Cray XC50 (ATERUI-II) and computers at Center for Computational Astrophysics, National Astronomical Observatory of Japan (CfCA/NAOJ). This work was supported by the Japan Society for the Promotion of Science (JSPS) Grant-in-Aid for Scientific Research (C) Grant Number 18K03711. This work was also supported by JSPS KAKENHI grant Nos. 18H01248 and 19H01933. DK acknowledges the support of the UK's Science \& Technology Facilities Council (STFC Grant ST/N000811/1).

\appendix
\section{Bar-driven gas inflow}

\subsection{Nuclear gas disc size}
\label{sec:gasring}

\begin{figure}
\begin{center}
\includegraphics[width=0.45\textwidth]{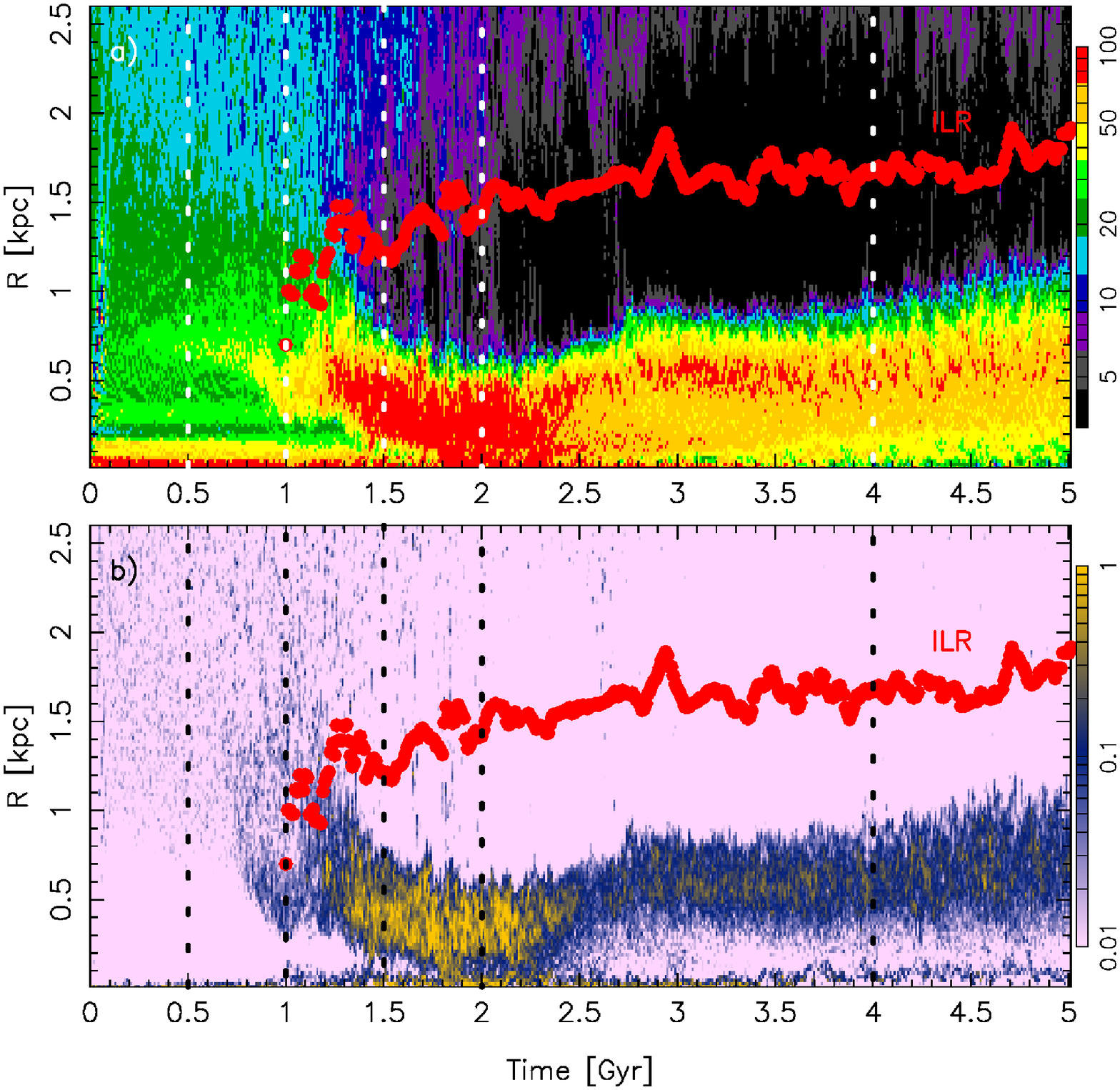}
\caption{
    (a) Temporal and radial variation of the azimuthally averaged gas surface density in the unit of $\rm M_\odot~pc^{-2}$ (logarithmic scale). The red thick dotted line marks the tempral variation of the location of the linear inner Lindblad resonance (ILR). The vertical dashed lines indicate the times corresponding to those of the snapshots shown in Fig.\ref{fig:SnapshotEvolution}.
    (b) Same of panel (a), but for the azimuthally averaged in-situ SFR density in the unit of $\rm M_\odot~yr^{-1}~kpc^{-2}$ (logarithmic scale).
}	
\label{fig:Ring}
\end{center}
\end{figure}

{
The formation mechanism of nuclear disc is still under debate. The widely believed theory is that the inner Lindblad resonance (ILR) is related to the location of nuclear disc \citep[][]{Combes1996,ButaCombes1996}. Alternatively, \citet{ReganTeuben2003,ReganTeuben2004} argued that the existence of $x_2$-orbits leads to the formation of nuclear disc. They also reminded that the ILR radius in strong bars is not a true resonance but an approximated radius where the $x_2$-orbits become stable \citep[][]{vanAlbadaSanders1982,ContopoulosGrosbol1989}.

Although the formation mechanism of nuclear gas disc in barred galaxies is beyond the scope of this study, we briefly present the evolution of the size of the nuclear gas disc in our simulation. Fig.~\ref{fig:Ring}(a) shows the temporal and radial variation of the gas surface density in the central region of the simulated galaxy. After the bar starts to grow at $t \simeq 1$ Gyr, the nuclear gas disc (i.e. the region with $\sim 10^2~\rm M_\odot~pc^{-2}$) forms around 0.5 kpc. The disc size continues to decrease until $t \simeq 2$ Gyr and then gradually increases. Such time evolution can be also seen in the middle row of Fig.\ref{fig:SnapshotEvolution}. The in-situ SFR density follows the same evolution as the surface density of the nuclear gas disc (Fig.~\ref{fig:Ring}(b)). For comparison, the time evolution of the ILR radius based on the liner perturbation theory is shown as the red dotted lines. We can see that the locations of the nuclear gas disc and star-forming disc are well inside of the ILR radius, although the time evolution of the disc sizes follow the similar evolutionary trend to the ILR radius. As \citet{Li+2015} suggested, the ILR radius seems to be related to constrain the nuclear disc size, but not the same as the size of the nuclear disc. 
}

\subsection{Bar-driven gas removal in the bar region}
\label{sec:gasdepletion}

\begin{figure}
\begin{center}
\includegraphics[width=0.45\textwidth]{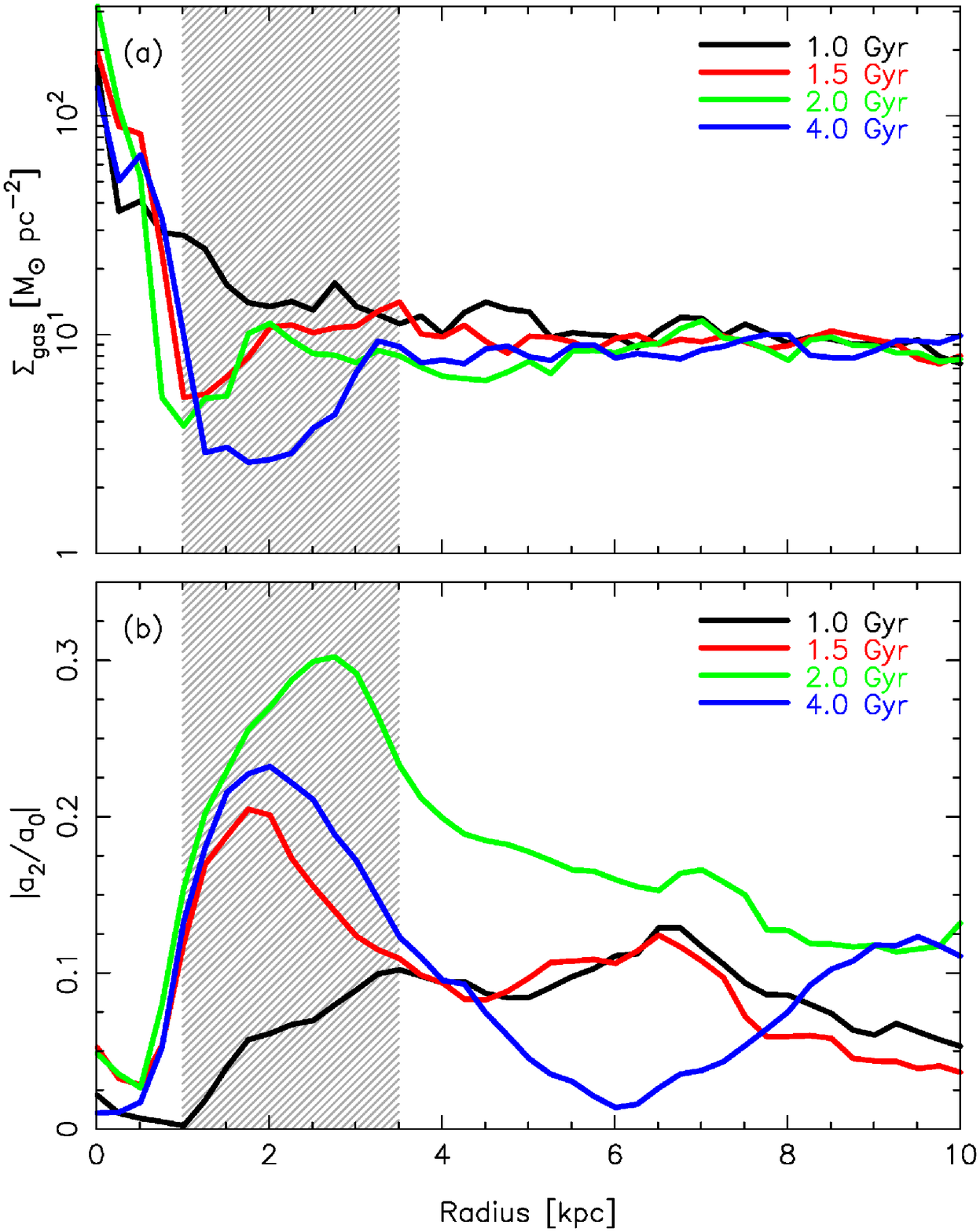}
\caption{
	Temporal change of radial profiles of (upper) gas surface density and (lower) normalized Fourier amplitude.
	Shaded area indicates the bar region. 
}	
\label{fig:ProfileEvolution}
\end{center}
\end{figure}

To understand the relationship between the bar growth and gas infall, we analysed the time evolution of the radial profiles of the bar amplitude \citep[$|a_2|$; e.g.][]{RixZaritsky1995} and gas surface density ($\Sigma_{\rm gas}$) as shown in Fig.~\ref{fig:ProfileEvolution}. Prior to the bar formation ($t = 1$ Gyr), the gas density followed an exponential profile as assumed in the initial condition. After the bar formation ($t > 1.5$ Gyr), the gas density started to decrease at the radius of around 1 kpc. At this point, the bar amplitude attained its maximum value and the gas density at the centre simultaneously started to increase. Subsequently, the gas-depleted region at $R \gtrsim 0.5$~kpc widened outwards and its outer edge reached a radius of around 3.3 kpc at $t = 4$ Gyr. The radius of the outer edge of the gas-depleted region was approximately equal to the ultra-harmonic (1:4) resonance radius (i.e. about 3.5 kpc) in our simulated barred galaxy. This depletion of the gas in the bar region leads to the reduction of star formation rate in the bar.

\bibliographystyle{mn2e}
\bibliography{ms}

\end{document}